\documentclass[12pt]{article}
\usepackage{amssymb,amsmath,epsfig}
\allowdisplaybreaks

\begin{document}
\title{\bf Minimally Deformed Regular Bardeen Black Hole Solutions in Rastall Theory}
\author{M. Sharif$^1$ \thanks{msharif.math@pu.edu.pk}~
and Malick Sallah$^{1,2}$ \thanks{malick.sallah@utg.edu.gm} \\
$^1$ Department of Mathematics and Statistics, The University of Lahore\\
1-KM Defence Road Lahore-54000, Pakistan.\\
$^2$ Department of Mathematics, The University of The Gambia,\\
Serrekunda, P.O. Box 3530, The Gambia.}
\date{}
\maketitle

\begin{abstract}
In this study, we utilize the minimal geometric deformation
technique of gravitational decoupling to extend the regular Bardeen
black hole, leading to the derivation of new black hole solutions
within the framework of Rastall theory. By decoupling the field
equations associated with an extended matter source into two
subsystems, we address the first subsystem using the metric
components of the regular Bardeen black hole. The second subsystem,
incorporating the effects of the additional source, is solved
through a constraint imposed by a linear equation of state. By
linearly combining the solutions of these subsystems, we obtain two
extended models. We then explore the distinct physical properties of
these models for specific values of the Rastall and decoupling
parameters. Our investigations encompass effective thermodynamic
variables such as density and anisotropic pressure, asymptotic
flatness, energy conditions, and thermodynamic properties including
Hawking temperature, entropy, and specific heat. The results reveal
that both models violate asymptotic flatness of the resulting
spacetimes. The violation of energy conditions indicate the presence
of exotic matter, for both models. Nonetheless, the energy density,
radial pressure, as well as the Hawking temperature exhibit
acceptable behavior, while the specific heat and Hessian matrix
suggest thermodynamic stability.
\end{abstract}

\section{Introduction}

The principle of minimal coupling, integral to general relativity
(GR), results in a divergence-free energy-momentum tensor
($\nabla_{\nu_1}T^{\nu_1}_{\nu_2}=0$) within curved spacetime. In
contrast, Rastall proposed a more expansive approach by discarding
the minimal coupling principle and introducing a nonminimal coupling
between matter and geometry instead \cite{1}. Rastall gravity posits
that this nonminimal coupling is linearly dependent on the Ricci
scalar $\mathcal{R}$, i.e.,
$\nabla_{\nu_1}T^{\nu_1}_{\nu_2}\propto\mathcal{R}_{,\nu_2}$. This
distinction becomes apparent in high-curvature, dense matter
environments, distinguishing it from GR, although it converges to GR
in vacuum conditions. Consequently, compact stars, such as neutron
stars, serve as prime candidates for testing Rastall gravity,
revealing unique characteristics when applied to such dense stellar
objects \cite{2}. Despite assertions that Rastall gravity is
essentially equivalent to GR \cite{3}, this claim has faced
criticism and been refuted due to misinterpretations of the matter
stress-energy tensor \cite{4}. Numerous studies across
thermodynamic, cosmological, and astrophysical domains have
demonstrated the non-equivalence of the two theories
\cite{5}-\cite{10}.

One of the strongest experimental confirmations of black holes comes
from images showing the shadow of a black hole, captured by the
Event Horizon Telescope Collaboration \cite{11,12}. Synge \cite{13}
is credited with the ground breaking work on the deflection of light
by highly gravitational stars. Bardeen \cite{14} determined that the
radius of the photon sphere for a Schwarzschild black hole is $3M$.
The discovery of black holes underscored the inadequacies of
Newtonian physics in explaining gravity and underscored the
significant implications of GR. Even without the presence of matter,
the Einstein field equations yield complex solutions like black
holes, which possess characteristics vastly different from those of
a flat Minkowski spacetime. The fascination with black holes arises
from the intricate interplay between classical and quantum physics,
which is essential for their understanding. Indeed, the classical
solutions to the Einstein field equations disclose both future
\cite{15} and past singularities \cite{16}. The interior of black
holes poses a conceptual challenge due to the existence of
singularities, typically concealed behind an event horizon
\cite{17}. A pivotal advancement in black hole physics was Stephen
Hawking pioneering work \cite{18}, which clarified the radiation
emission from black hole event horizons. This remarkable discovery
transformed black holes into crucial experimental sites, providing a
unique laboratory for investigating the complex aspects of
gravitational theories.

The renowned singularity theorem by Penrose \cite{15} asserts that
under specific conditions, singularities are unavoidable in GR. This
aligns with the fact that the earliest known exact black hole
solutions in GR, feature a singularity within the event horizon.
Despite this, it is widely believed that singularities are not
physical entities but are instead artifacts produced by classical
gravitational theories, and they do not actually exist in nature.
Quantum arguments proposed by Sakharov \cite{19} and Gliner
\cite{20} suggest that spacetime singularities could be circumvented
by matter sources with a de Sitter core at the center of the
spacetime. Building on this concept, Bardeen introduced the first
static spherically symmetric regular black hole solution \cite{21}.
This model was inspired by the Reissner-Nordstrom spacetime. It
describes a standard black hole that adheres to the weak energy
condition, significantly influencing the trajectory of studies
regarding the presence or prevention of singularities. Subsequent
models of regular black holes which demonstrate violations of the
strong energy condition have since been proposed
\cite{22}-\cite{27}, thereby challenging the singularity theorems.
Ayon-Beato and Garcia \cite{28} established that these models could
be viewed as the gravitational fields of nonlinear electric or
magnetic monopoles, suggesting that nonlinear electromagnetic fields
might be the physical sources of regular black holes. This
interpretation is supported by other researchers in the field as
well \cite{29}.

Exploring the thermodynamic characteristics of black holes is
essential for advancing our knowledge of fundamental physics. These
characteristics, including temperature, entropy, and specific heat,
serve as a bridge connecting quantum mechanics, GR, and statistical
mechanics. Investigating how black holes interact with radiation
provides researchers with valuable insights into the nature of
spacetime, the behavior of quantum fields in intense gravitational
environments, and the underlying principles that dictate the
universe ultimate fate. This interdisciplinary approach not only
enhances our understanding of black holes but also offers a deeper
grasp of the fundamental laws of physics under extreme conditions.
Bekenstein \cite{30} first linked black hole surface area to
entropy, and subsequently, Hawking \cite{18} showed that black holes
with surface gravity (denoted as $k$) radiate at a temperature of
($\frac{k}{2\pi}$). However, the Bekenstein-Hawking radiation
introduces the information loss paradox due to thermal evaporation.
To tackle this paradox, Hawking and colleagues \cite{31} recently
proposed the concept of soft hair. Owing to the pivotal
contributions of Bekenstein and Hawking to black hole
thermodynamics, black hole radiation has garnered significant
interest from researchers. Recently, there has been a notable
increase in the study of black holes and their thermodynamic
properties within the context of Rastall gravity theory
\cite{32}-\cite{36}.

As is well known, solving the Einstein field equations is
particularly challenging, especially in scenarios involving
spherical symmetry. A recent breakthrough that addresses this
complexity is the introduction of a new method called gravitational
decoupling via the minimal geometric deformation (MGD) scheme
\cite{37}. It later developed into a gravitational source decoupling
scheme, enabling the extension of isotropic spherical solutions of
the Einstein field equations to anisotropic contexts \cite{38}. Over
time, this method has gained widespread acceptance across various
modified theories of gravity \cite{39}-\cite{44} including the
Rastall theory \cite{45}-\cite{47}, significantly contributing to
the development of new solutions for the Einstein equations and
their extensions. The MGD and the extended geometric deformation
(EGD) are the two techniques that comprise the gravitational
decoupling scheme. The main difference is that MGD adjusts only the
radial part of the spacetime metric, whereas EGD changes both the
temporal and radial components. Additionally, MGD is limited to
situations where decoupled sources interact solely through gravity,
and it cannot be applied to cases with energy exchange between
sources. It is worthy to mention, however, that these deformation
schemes do not alter the spherical symmetry of the spacetime
configuration.

In this study, we profit from the MGD scheme to generalize the
regular Bardeen black hole in the context of Rastall theory. We thus
obtain new extended solutions, which are analyzed and compared with
existing results. The outline of this paper is as follows. In
Section \textbf{2}, we present the Rastall field equations for a
dual matter source and apply the MGD scheme to these equations.
Section \textbf{3} pertains to the derivation and analysis of two
extended solutions, with emphasis on such physical features as
asymptotic flatness, and energy conditions. Section \textbf{4}
entails a thermodynamic analysis of the generalized solutions
obtained. Finally, we summarize and analyze our results in Section
\textbf{5}.

\section{Rastall Field Equations}

In the context of curved spacetime, Rastall theory \cite{1} diverges
from the conventional approach by rejecting the idea of a
divergence-free stress-energy tensor, i.e., $\nabla_\upsilon
T^{\upsilon\omega}\neq 0$. This theory introduces a non-minimal
interaction between geometry and matter, achieved by allowing the
stress-energy tensor to have a non-zero divergence, i.e.,
\begin{equation}\label{1}
\nabla^\omega T^R_{\upsilon\omega}=\frac{\xi}{4}
g_{\upsilon\omega}\nabla^\omega\mathcal{R},
\end{equation}
where $\xi$ and $\mathcal{R}$ denote the Rastall parameter and Ricci
scalar, respectively. Thus by this equation, Rastall posited that
the covariant divergence of the stress-energy tensor is proportional
to the divergence of the curvature scalar, $\mathcal{R}$, with the
Rastall parameter as the proportionality constant. Clearly, the
usual conservation result of GR is regained in the event $\xi\mapsto
0$ or $\mathcal{R}\mapsto 0$ (flat spacetime). The degree of
deviation of the Rastall theory from GR is thus specified by the
Rastall parameter, which encapsulates the nature of the non-minimal
coupling between geometry and matter. The stress-energy tensor
defined in Eq.\eqref{1} satisfies the Rastall field equations
\cite{3}
\begin{equation}\label{2}
\mathcal{R}_{\upsilon\omega}-\frac{1}{2}\mathcal{R}g_{\upsilon\omega}
+\frac{\xi}{4}\mathcal{R}g_{\upsilon\omega}=\kappa
T^R_{\upsilon\omega},
\end{equation}
where $\mathcal{R}_{\upsilon\omega},~g_{\upsilon\omega},$ and
$\kappa$ denote the Ricci tensor, metric tensor and coupling
constant, respectively. Again, we observe that these field equations
reduce to the field equations of GR in the event the Rastall
parameter vanishes (i.e., $\xi\mapsto 0$). The field equations
\eqref{2} can be rewritten as
\begin{equation}\label{3}
\mathcal{R}_{\upsilon\omega}-\frac{1}{2}\mathcal{R}g_{\upsilon\omega}
=\kappa \hat{T}_{\upsilon\omega},
\end{equation}
where
\begin{equation}\label{4}
\hat{T}_{\upsilon\omega}=T^R_{\upsilon\omega}-\frac{\xi}{4(\xi-1)}T^R
g_{\upsilon\omega}.
\end{equation}
From Eq.\eqref{4} above, we obtain the following explicit expression
for the Rastall stress-energy tensor
\begin{equation}\label{5}
T^R_{\upsilon\omega}=\hat{T}_{\upsilon\omega}-\frac{\xi}{4}\hat{T}g_{\upsilon\omega},
\end{equation}
where
\begin{equation}\label{6}
\hat{T}_{\upsilon\omega}=(\rho+P_t)V_\upsilon
V_\omega-P_tg_{\upsilon\omega}+(P_t-P_r)Y_\upsilon Y_\omega,
\end{equation}
is identified as the anisotropic energy-momentum tensor, and
$\hat{T}$ as its trace. Here, $V_\upsilon=(\sqrt{g_{00}},0,0,0)$ and
$Y_\upsilon=(0,-\sqrt{-g_{11}},0,0)$ denote the 4-vector and
4-velocity, respectively, and satisfy the relations
\begin{equation}\nonumber
V^\upsilon Y_\upsilon=0,\quad V^\upsilon V_\upsilon=1,\quad
Y^\upsilon Y_\upsilon=-1.
\end{equation}

As we seek to extend the regular Bardeen black hole solution in this
study, we consider a modification of the field equations \eqref{1}
wherein an extra matter source is gravitationally coupled to the
seed source which is specified by the metric potentials of the
aforementioned solution. The modified field equations are thus given
by
\begin{equation}\label{7}
\mathcal{R}_{\upsilon\omega}-\frac{1}{2}\mathcal{R}g_{\upsilon\omega}
+\frac{\xi}{4}\mathcal{R}g_{\upsilon\omega}=\kappa{T}^{(Tot)}_{\upsilon\omega},
\end{equation}
where
\begin{equation}\label{8}
{T}^{(Tot)}_{\upsilon\omega}=T^R_{\upsilon\omega}+\sigma\chi_{\upsilon\omega},
\end{equation}
and $T^R_{\upsilon\omega}$ is given by Eq.\eqref{5}. In this
context, $\chi_{\upsilon\omega}$ represents a supplementary source
gravitationally linked to the primary source $T^R_{\upsilon\omega}$
through the decoupling parameter $\sigma$. This supplementary source
induces anisotropies within self-gravitating structures and may
introduce new fields defined by vectors, tensors, and scalars. The
geometry of the spacetime is described by the metric
\begin{equation}\label{9}
ds^2=e^{\eta_1(r)}dt^2-e^{\eta_2(r)}dr^2-r^2(d\theta^2+\sin^2\theta
d\phi^2).
\end{equation}
With this metric, the field equations \eqref{7} become
\begin{align}\nonumber
\kappa\left[\rho-\frac{\xi}{4}(\rho-P_r-2P_t)+\sigma\chi^0_0
\right]&=\frac{1}{r^2}+e^{-\eta_2}\bigg(\frac{\eta_2^\prime}{r}
-\frac{1}{r^2}\bigg)+\frac{\xi e^{-\mu_2}}{4}
\bigg(\eta_1^{\prime\prime}+\frac{\eta_1^\prime(\eta_1^\prime
-\eta_2^\prime)}{2}\bigg)\\\label{10} &+\frac{\xi
e^{-\eta_2}}{4}\bigg(\frac{2(\eta_1^\prime -\eta_2^\prime)}{r}
+\frac{2}{r^2}\bigg)-\frac{\xi}{2r^2},
\end{align}
\begin{align}\nonumber
\kappa\left[P_r+\frac{\xi}{4}(\rho-P_r-2P_t)-\sigma\chi^1_1\right]&=
-\frac{1}{r^2}+e^{-\eta_2}\bigg(\frac{\eta_1^\prime}{r}+\frac{1}{r^2}\bigg)
-\frac{\xi e^{-\eta_2}}{4}\bigg(\eta_1^{\prime\prime}
+\frac{\eta_1^\prime(\eta_1^\prime-\eta_2^\prime)}{2}\bigg)\\\label{11}
&-\frac{\xi e^{-\eta_2}}{4}\bigg(\frac{2(\eta_1^\prime
-\eta_2^\prime)}{r} +\frac{2}{r^2}\bigg)+\frac{\xi}{2r^2},
\end{align}
\begin{align}\nonumber
\kappa\left[P_t+\frac{\xi}{4}(\rho-P_r-2P_t)-\sigma\chi^2_2\right]&=e^{-\eta_2}
\bigg(\frac{\eta_1^{\prime\prime}}{2}+\frac{\eta_1^{\prime^2}}{4}
-\frac{\eta_1^\prime\eta_2^\prime}{4}+\frac{\eta_1^\prime}{2r}
-\frac{\eta_2^\prime}{2r}\bigg)+\frac{\xi}{2r^2}\\\label{12}
&-\frac{\xi e^{-\eta_2}}{4}\bigg(\eta_1^{\prime\prime}
+\frac{\eta_1^\prime(\eta_1^\prime-\eta_2^\prime)}{2}
+\frac{2(\eta_1^\prime-\eta_2^\prime)}{r}+\frac{2}{r^2}\bigg).
\end{align}
With respect to this system, the total energy-momentum tensor
defined in Eq.\eqref{8} is conserved as
\begin{equation}\label{13}
\nabla^\omega T^{(Tot)}_{\upsilon\omega}=\frac{dP_r}{dr}
+\frac{\eta_1^\prime}{2}(\rho+P_r)+\sigma\frac{\eta_1^\prime}{2}
(\chi_1^1-\chi_0^0)+\frac{2}{r}(P_r-P_t)\\+\sigma\frac{d\chi_1^1}{dr}
+\frac{2\sigma}{r}(\chi^1_1-\chi^2_2)=0.
\end{equation}
The system of equations \eqref{10}-\eqref{12} constitutes three
non-linear ordinary differential equations in eight unknowns, given
by $\rho,~P_r,~P_t,~\eta_1,~\eta_2,~\chi^0_0,~\chi_1^1,~\chi^2_2$.
Additionally, the prime notation denotes differentiation with
respect to $r$. From this system, we infer the following effective
variables
\begin{equation}\label{14}
\tilde{\rho}=\rho+\sigma\chi^0_0,\quad
\tilde{P_r}=P_r-\sigma\chi^1_1,\quad \tilde{P_t}=P_t-\sigma\chi^2_2.
\end{equation}
These effective variables prompt an anisotropy defined as
\begin{equation}\label{15}
\digamma=\tilde{P_t}-\tilde{P_r}=(P_t-P_r)+\sigma(\chi_1^1-\chi^2_2).
\end{equation}

In what follows, we exploit the gravitational decoupling technique
via the MGD scheme to decouple the field equations
\eqref{10}-\eqref{12}. The decoupling process splits the field
equations into two sets, the first of which corresponds to the seed
source and will be specified by by the metric components of the
regular Bardeen black hole solution \cite{21}. The second set
describes the effects of the extra source and will be solved by
employing appropriate constraints. Using Eq.\eqref{14}, we obtain a
linear combination of the solutions of the subfield equations,
constituting a solution of the system of the field equations
\eqref{10}-\eqref{12}.

\section{A Gravitational Decoupling Scheme}

The field equations become increasingly complex with the addition of
a second source to the original anisotropic fluid, introducing more
unknown variables. To achieve an exact solution, it is necessary to
limit the degrees of freedom and choose a specific strategy or set
of constraints. Thus, we employ a systematic technique known as
gravitational decoupling, which when applied to the field equations,
allows for the derivation of a solution. A fascinating feature of
this technique is that it transforms the temporal-radial metric
potentials into a new reference frame, thereby simplifying the
equations. However, we proceed with the MGD scheme of gravitational
decoupling wherein only the radial component of the metric is
altered. This implies that the temporal metric component is
preserved under the MGD scheme. To proceed, we set the coupling
parameter $\sigma$ to zero and examine an anisotropic solution
$(\eta_3,~\eta_4,\rho,~P_r,~P_t)$ of the field equations
\eqref{10}-\eqref{12}, described by the metric
\begin{equation}\label{16}
ds^2=e^{\eta_3(r)}dt^2-\frac{1}{\eta_4(r)}dr^2
-r^2(d\theta^2+\sin^2\theta d\phi^2),
\end{equation}
where
\begin{equation}\label{17}
\eta_4(r)=1-\frac{2m(r)}{r},
\end{equation}
with $m$ denoting the Misner-Sharp mass function.

The geometric deformation on the metric functions are applied
through the following linear transformations
\begin{equation}\label{18}
\eta_3(r)\mapsto\eta_1(r)=\eta_3(r)+\sigma
f(r),\quad\eta_4(r)\mapsto e^{-\eta_2(r)}=\eta_4(r)+\sigma h(r),
\end{equation}
with $f(r)$ and $h(r)$ as the deformations on $g_{tt}$ and $g_{rr}$,
respectively. Owing to the MGD scheme, we set $f(r)=0$, implying
that the temporal component of the metric is preserved. Substituting
these transformations into the field equations, we obtain two sets
of subfield equations. The first of these corresponds to $\sigma=0$
and is given by
\begin{align}\nonumber
\kappa\left[\rho-\frac{\xi}{4}(\rho-P_r-2P_t)\right]&=\eta_4\bigg(
\frac{\xi\eta_3^{\prime\prime}}{4}-\frac{1}{r^2}+\frac{\xi\eta_3^{
\prime^2}}{8}+\frac{\xi\eta_3^\prime}{2r}+\frac{\xi}{2r^2}\bigg)
\\\label{19}&+\eta_4^\prime\bigg(\frac{\xi}{2r}
+\frac{\xi\eta_3^\prime}{8}-\frac{1}{r}\bigg)-\frac{\xi}{2r^2}+
\frac{1}{r^2},\\\nonumber\kappa\left[P_r+\frac{\xi}{4}(\rho-P_r-2P_t)
\right]&=\eta_4\bigg(\frac{\eta_3^\prime}{r}-\frac{\xi\eta_3^{\prime\prime}}{4}
+\frac{1}{r^2}-\frac{\xi\eta_3^{\prime^2}}{8}
-\frac{\xi\eta_3^\prime}{2r}-\frac{\xi}{2r^2}\bigg)\\\label{20}
&-\eta_4^\prime\bigg(\frac{\xi\eta_3^\prime}{8}
+\frac{\xi}{2r}\bigg)+\frac{\xi}{2r^2}-\frac{1}{r^2},
\\\nonumber\kappa\left[P_t+\frac{\xi}{4}(\rho-P_r-2P_t)\right]&
=\eta_4\bigg(\frac{\eta_3^{\prime\prime}}{2}
+\frac{\eta_3^{\prime^2}}{4}+\frac{\eta_3^\prime}{2r}
-\frac{\xi\eta_3^{\prime\prime}}{4} -\frac{\xi\eta_3^{\prime^2}}{8}
-\frac{\xi\eta_3^\prime}{2r}-\frac{\xi}{2r^2}\bigg)\\\label{21}&
+\eta_4^\prime\bigg(\frac{\eta_3^\prime}{4}
+\frac{1}{2r}-\frac{\xi\eta_3^\prime}{8}
-\frac{\xi}{2r}\bigg)+\frac{\xi}{2r^2}.
\end{align}
The conservation equation with respect to this set becomes
\begin{equation}\label{22}
\frac{dP_r}{dr}+\frac{\eta_3^\prime}{2}(\rho+P_r)+
\frac{2}{r}(P_r-P_t)=0.
\end{equation}
It can be observed that the system \eqref{19}-\eqref{21}, consists
of three equations in the five variables
$\big(\rho,~P_r,~P_t,~\eta_3,~\eta_4\big)$. It therefore suffices to
adopt two constraints to close this system. For this, we shall
employ the metric potentials of the regular Bardeen black hole
\cite{21}.

The second set incorporates the effects of the extra source
$\chi_{\upsilon\omega}$, and is obtained by turning on the effect of
the decoupling parameter $\sigma$. This set is given by the system
\begin{align}\label{23}
\kappa\chi^0_0&=h\bigg(\frac{\xi}{2r^2}
+\frac{\xi\eta_3^\prime}{2r}+\frac{\xi\eta_3^{\prime^2}}{8}
+\frac{\xi\eta_3^{\prime\prime}}{4}-\frac{1}{r^2}\bigg)
+h^\prime\bigg(\frac{\xi}{2r}+\frac{\xi\eta_3^\prime}{8}
-\frac{1}{r}\bigg),
\\\label{24}
\kappa\chi^1_1&=h\bigg(\frac{\xi}{2r^2}
+\frac{\xi\eta_3^\prime}{2r}+\frac{\xi\eta_3^{\prime^2}}{8}
+\frac{\xi\eta_3^{\prime\prime}}{4}
-\frac{1}{r^2}-\frac{\eta_3^\prime}{r}\bigg)
+h^\prime\bigg(\frac{\xi}{2r} +\frac{\xi\eta_3^\prime}{8}\bigg),
\\\nonumber \kappa\chi^2_2&=h\bigg(\frac{\xi}{2r^2}
+\frac{\xi\eta_3^\prime}{2r}+\frac{\xi\eta_3^{\prime^2}}{8}
+\frac{\xi\eta_3^{\prime\prime}}{4}-\frac{\eta_3^\prime}{2r}
-\frac{\eta_3^{\prime^2}}{4}-\frac{\eta_3^{\prime\prime}}{2}\bigg)
\\\label{25}&+h^\prime\bigg(\frac{\xi}{2r}
+\frac{\xi\eta_3^\prime}{8}-\frac{1}{2r}
-\frac{\eta_3^\prime}{4}\bigg).
\end{align}
This system is conserved according to the equation below
\begin{equation}\label{26}
\frac{d\chi^1_1}{dr}+\frac{\eta_3^\prime}{2}(\chi_1^1-\chi^0_0)
+\frac{2}{r}(\chi_1^1-\chi_0^0)=0.
\end{equation}
It is observed that this system comprises three equations in four
variables, viz $(\chi_0^0,~\chi_1^1,~\chi_2^2,~h)$. A single
constraint is thus sufficient to close this system. It is worthy to
highlight that the subfield equations \eqref{19}-\eqref{21} and
\eqref{23}-\eqref{25} are each individually conserved. This implies
that there is a null exchange of energy momentum between the
sources, which is a necessary condition for the applicability of the
MGD scheme.

\section{Extending the Regular Bardeen Black Hole Solution}

The regular Bardeen black hole solution is described by the line
element \cite{21}
\begin{equation}\label{27}
ds^2=\bigg(1-\frac{2Mr^2}{(r^2+e^2)^{\frac{3}{2}}}\bigg)dt^2-
\bigg(1-\frac{2Mr^2}{(r^2+e^2)^{\frac{3}{2}}}\bigg)^{-1}dr^2
-r^2(d\theta^2+\sin^2\theta d\phi^2),
\end{equation}
where $M$ and $e$ denote the mass and magnetic monopole charge of
the black hole, respectively. The metric above has a coinciding
Killing ($r_H$) and causal horizons ($r_h$) at the surface. The
coincidence of the Killing and causal horizons is a necessary
condition for the deformed Bardeen metric \eqref{29} to denote a
proper black hole. These horizons are specified by the conditions
$e^{\eta_1}=0$ and $e^{-\eta_2}=0$, respectively \cite{48}. The
Killing horizon for the regular Bardeen black hole metric is thus
obtained as
\begin{equation}\label{28}
r_H=\sqrt{Ae^2+\frac{B+C}{3}+\frac{4M^2}{3}},
\end{equation}
where
\begin{align}\nonumber
A&=-\frac{2^{8/3} M^2}{\sqrt[3]{27 e^4 M^2-72 e^2 M^4+3 \sqrt{81 e^8
M^4-48 e^6 M^6}+32 M^6}}-1,\\\nonumber
\\\nonumber B&=\frac{
2^{11/3}M^4}{\sqrt[3]{27 e^4 M^2-72 e^2 M^4+3 \sqrt{81 e^8 M^4-48
e^6 M^6}+32 M^6}},\\\nonumber
\\\nonumber C&=\sqrt[3]{54 e^4 M^2-144 e^2 M^4+6
\sqrt{81 e^8 M^4-48 e^6 M^6}+64 M^6}.
\end{align}
Using the minimal deformation of the metric \eqref{27}, we obtain
extensions of the regular Bardeen black hole solution. The line
element for the minimally deformed Bardeen black hole solution is
given by
\begin{equation}\label{29}
ds^2=\bigg(1-\frac{2Mr^2}{(r^2+e^2)^{\frac{3}{2}}}\bigg)dt^2-
\bigg(1-\frac{2Mr^2}{(r^2+e^2)^{\frac{3}{2}}}+\sigma
h(r)\bigg)^{-1}dr^2 -r^2(d\theta^2+\sin^2\theta d\phi^2),
\end{equation}
where $h(r)$ is obtained from the system \eqref{23}-\eqref{25}, by
specifying a constraint on the extra source,
$\chi_{\upsilon\omega}$. These constraints are specified by the
linear equation of state (EoS) \cite{48}
\begin{equation}\label{30}
\chi_0^0+\lambda\chi^1_1+\tau\chi^2_2=0,
\end{equation}
where $\lambda$ and $\tau$ are real constants.

In what follows, we obtain two extensions of the regular Bardeen
black hole solution, using two specific cases of the EoS mentioned
above \eqref{30}.

\subsection{Model I: Traceless Additional Source}

The additional source $\chi_{\upsilon\omega}$ is termed as
conformally symmetric when its energy-momentum tensor possesses a
null trace. As $\chi^2_2=\chi^3_3$ (due to the spherical symmetry),
a traceless additional source is implied if
\begin{equation}\label{31}
\chi_0^0+\chi^1_1+2\chi^2_2=0,
\end{equation}
i.e., when $\lambda=1$ and $\tau=2$ in \eqref{30}. Using the system
\eqref{23}-\eqref{25}, Eq.\eqref{31} above simplifies to
\begin{equation}\label{32}
\frac{(\xi -1) \left(r h'(r) \left(r \eta_3'(r)+4\right)+h(r)
\left(2 r^2 \eta_3''(r)+\left(r
\eta_3'(r)+2\right)^2\right)\right)}{2 r^2}=0,
\end{equation}
from which $h(r)$ can be obtained. Due to a complicated expression,
the result for $h(r)$ from \eqref{32} is not written. Nonetheless,
its graph is shown in Figure \textbf{1}. We would like to mention
here that for all the plots, we have used values of the Rastall and
decoupling parameters as $\xi=0.2,0.6$ and
$\sigma=0.2,0.4,0.6,0.8,1$, respectively, while the magnetic
monopole charge has been fixed at $e=1$. Additionally, we have used
$M=1$, in order to cater for a region accessible to an outer
observer. Substituting the deformation function obtained from
Eq.\eqref{32} above into the deformed Bardeen metric \eqref{29}, we
obtain an extended solution. This extended model is described by the
following effective variables
\begin{align}\nonumber
\tilde{\rho}&=\frac{3 e^2 M \left(4 e^2 (\xi -1)-(\xi +4)
r^2\right)}{16 \pi  (\xi -1)
\left(e^2+r^2\right)^{7/2}}+\sigma\chi_0^0,\\\nonumber\tilde{P_r}&=\frac{3
e^2 M \left((\xi +4) r^2-4 e^2 (\xi -1)\right)}{16 \pi (\xi -1)
\left(e^2+r^2\right)^{7/2}}-\sigma\chi_1^1,
\\\label{33}\tilde{P_t}&=-\frac{3 e^2 M \left(4 e^2 (\xi -1)+(6-11 \xi )
r^2\right)}{16 \pi  (\xi -1)
\left(e^2+r^2\right)^{7/2}}-\sigma\chi_2^2.
\end{align}
\begin{figure}\center
\epsfig{file=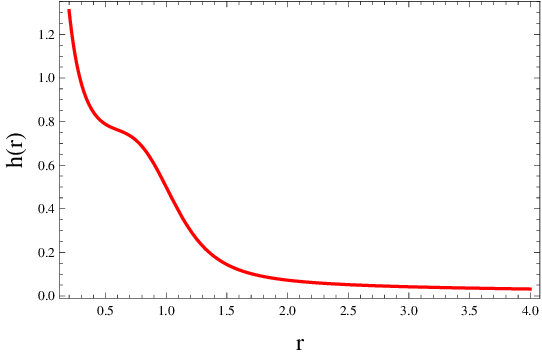,width=0.475\linewidth}
\epsfig{file=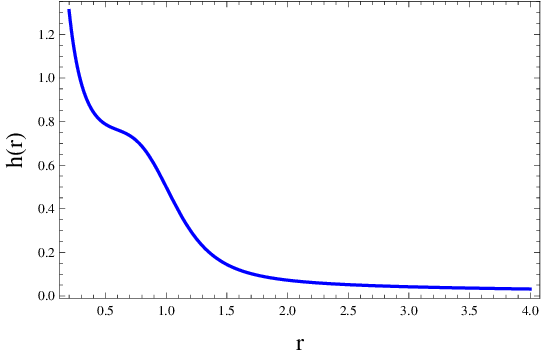,width=0.475\linewidth}\caption{Graphs of
deformation function $h(r)$ against $r$ with $\xi=0.2$ (left), $0.6$
(right) for model I.}
\end{figure}

It is fundamental to investigate whether the extended model
preserves the regularity of the Bardeen black hole. It is observed
that the inclusion of the deformation function $h(r)$ creates the
divergence of the extended models \eqref{29}, from the standard
Bardeen black hole \eqref{27}. This implies that the extended model
is regular if the associated deformation function is nonsingular at
the core. Figure \textbf{1} shows that the deformation function
obtained for the first model is regular at the core, thus implying
the regularity of the associated extended model. We additionally
provide a graph of the modified metric coefficient, enabling us to
evaluate the asymptotic flatness of the new spacetime. If the metric
potentials of a spacetime converge to 1 as the radial distance
increases arbitrarily, the spacetime is said to be asymptotically
flat. The gravitational field gradually decreases in such a
spacetime and vanishes completely at great distances from a huge
entity. This suggests that spacetime seems flat at large distances,
much like the flat spacetime that special relativity describes, when
gravity is essentially nonexistent.
\begin{figure}\center
\epsfig{file=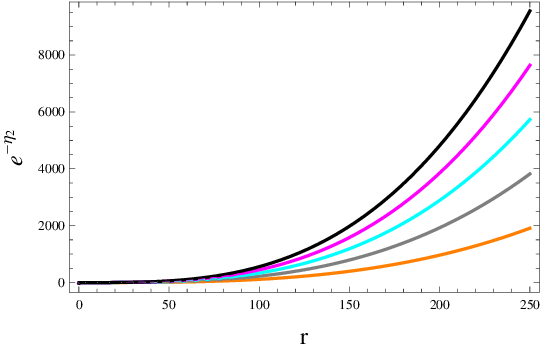,width=0.5\linewidth}\caption{Graph of $e^{-\eta_2}$
against $r$ with $\xi=0.2$ (solid), $0.6$ (dashed), $\sigma=0.2$
(orange), $0.4$ (gray), $0.6$ (cyan), $0.8$ (magenta) and $1$
(black) for model I.}
\end{figure}
\begin{figure}\center
\epsfig{file=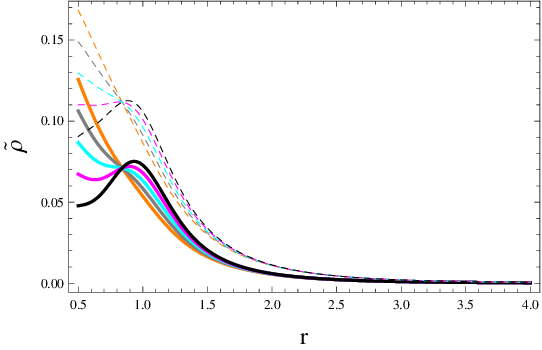,width=0.475\linewidth}
\epsfig{file=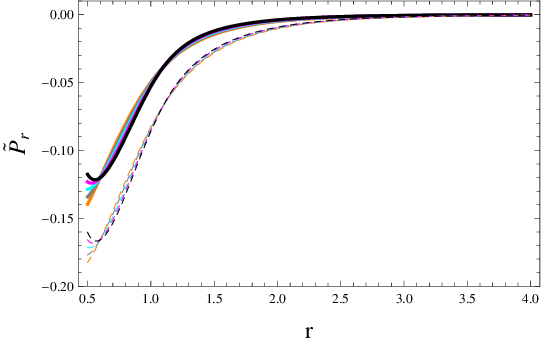,width=0.475\linewidth}
\epsfig{file=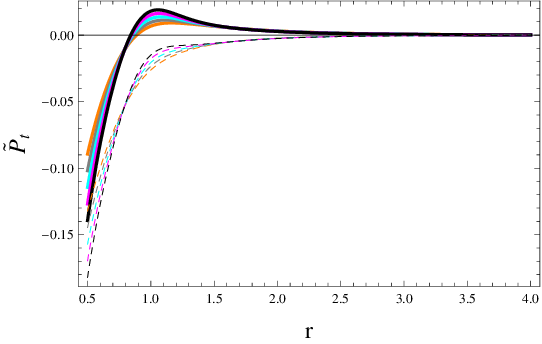,width=0.475\linewidth}\caption{Graphs of effective
parameters against $r$ with $\xi=0.2$ (solid), $0.6$ (dashed),
$\sigma=0.2$ (orange), $0.4$ (gray), $0.6$ (cyan), $0.8$ (magenta)
and $1$ (black) for model I.}
\end{figure}

Figure \textbf{2} portrays that the criteria for asymptotic flatness
has been violated by the radial metric component. It should be noted
that we only employed the graphical analysis for the deformed radial
metric component of the extended solution in our analysis of the
asymptotic flatness. This is because it is readily inferred from the
deformed metric \eqref{29}, that $\lim_{r\mapsto\infty}g_{tt}=1$.
The effective parameters \eqref{33} are plotted in Figure
\textbf{3}. We observe a positive density and negative radial
pressure, which is the acceptable behavior for these parameters. A
negative radial pressure implies an inward force that amplifies the
gravitational pull of a black hole. This idea fits well with the
current understanding, where matter collapses into a singularity due
to immense gravitational forces. In theoretical models, negative
radial pressure frequently aids in explaining various phenomena,
such as the accelerated expansion of the universe, as observed in
theories involving dark energy with negative pressure. Additionally,
the energy density and radial pressure exhibit a direct and indirect
variance, respectively, to the Rastall parameter $\xi$. The
tangential pressure however, assumes both negative and positive
values in its domain whilst varying inversely with the Rastall
parameter. With respect to the decoupling parameter $\sigma$, the
effective parameters exhibit a discrepancy as they vary both
directly and indirectly in different intervals within their domains.

To ascertain the characteristics of the matter source,
$T^{(Tot)}_{\upsilon\omega}$, we additionally plot the energy
conditions. The energy-momentum tensor of the source is constrained
by these energy conditions, which comprise dominant, strong, null,
and weak classifications. The source is regarded as normal if the
energy criteria are met, and exotic if certain energy requirements
are not met. The classification of these energy conditions are given
as follows
\begin{itemize}
\item Dominant Energy Conditions\\
$\tilde{\rho}\geq |\tilde{P_r}|,\quad\tilde{\rho}\geq
|\tilde{P_t}|.$
\item Strong Energy Conditions\\
$\tilde{\rho}\geq-\tilde{P_r},\quad\tilde{\rho}\geq-\tilde{P_t},
\quad\tilde{\rho}+\tilde{P_r}\geq-2\tilde{P_t}.$
\item Null Energy Conditions\\
$\tilde{\rho}\geq-\tilde{P_r},\quad\tilde{\rho}\geq-\tilde{P_t}.$
\item Weak Energy Conditions\\
$\tilde{\rho}\geq 0,\quad\tilde{\rho}\geq-\tilde{P_r},
\quad\tilde{\rho}\geq-\tilde{P_t}.$
\end{itemize}
These graphs of the energy conditions plotted in Figure \textbf{4},
depict that the matter source $T^{(Tot)}_{\upsilon\omega}$ is
exotic.
\begin{figure}\center
\epsfig{file=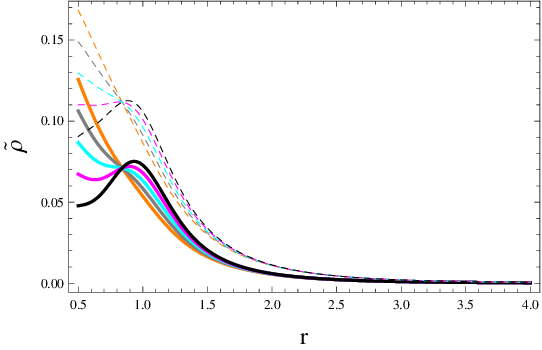,width=0.475\linewidth}
\epsfig{file=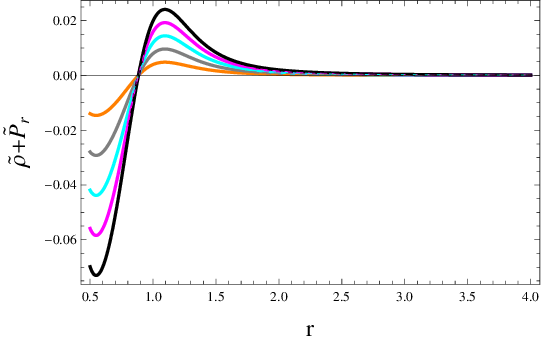,width=0.475\linewidth}
\epsfig{file=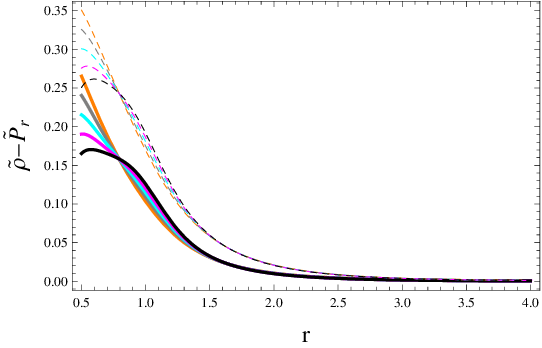,width=0.475\linewidth}
\epsfig{file=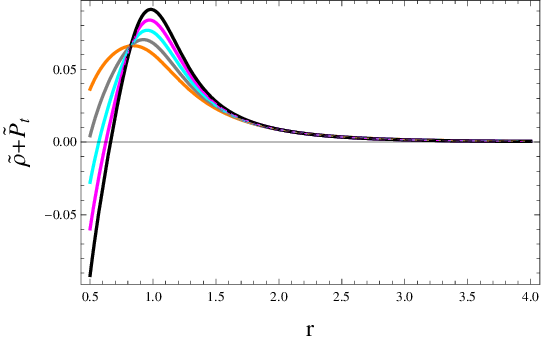,width=0.475\linewidth}
\epsfig{file=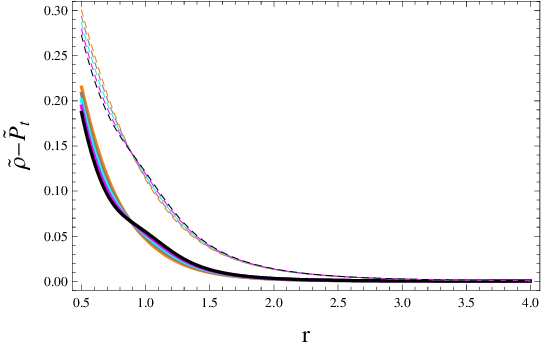,width=0.475\linewidth}
\epsfig{file=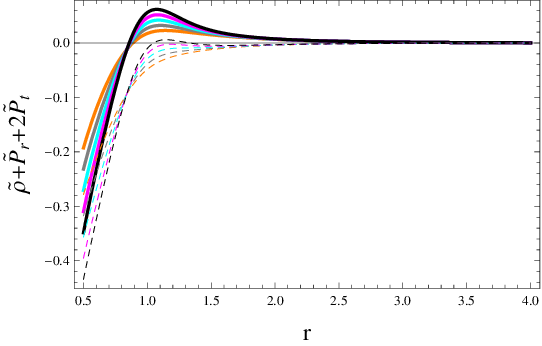,width=0.475\linewidth} \caption{Graphs of energy
conditions against $r$ with $\xi=0.2$ (solid), $0.6$ (dashed),
$\sigma=0.2$ (orange), $0.4$ (gray), $0.6$ (cyan), $0.8$ (magenta)
and $1$ (black) for model I.}
\end{figure}

\subsection{Model II: A Barotropic EoS}

A barotropic EoS is a unique case of a polytropic EoS. The
additional source $\chi_{\upsilon\omega}$ describes a barotropic
fluid if it satisfies the EoS
\begin{equation}\label{34}
\delta(\chi^0_0)-\chi^1_1=0,
\end{equation}
where $\delta$ denotes a nonnegative parameter which incorporates
information about the temperature. It can be observed that this
equation is a special case of the linear EoS \eqref{30} with
$\lambda=-\frac{1}{\delta}$ and $\tau=0$. Using the system
\eqref{23}-\eqref{25}, the barotropic EoS \eqref{34} becomes
\begin{align}\nonumber
\frac{r h'(r)(\delta+1)\xi  r \eta_3'(r)}{8r^2}&+\frac{r
h(r)\eta_3'(r) \left((\delta-1)\xi r\eta_3'(r)+4 (\delta-1) \xi
+8\right)}{8r^2}\\\label{35}+\frac{4(\delta(\xi
-2)+\xi)}{8r^2}&+\frac{r h(r)2 (\delta-1)\xi
r\eta_3''(r)}{8r^2}+\frac{4h(r) (\delta-1)(\xi -2)}{8r^2}=0,
\end{align}
from which we obtain the deformation function $h(r)$. Due to the
complexity and length of the expression, the explicit form of $h(r)$
is omitted. Instead, its graph is shown below. By substituting this
deformation function into the minimally deformed Bardeen metric from
Eq.\eqref{29}, we derive another extended solution. This new model
is defined by the effective parameters outlined in Eq.(33). The
distinction here is that the deformation function used is derived
from Eq.\eqref{35}. The deformation function in Figure \textbf{5}
indicates the absence of singularities within its domain.
Consequently, by the same reasoning as in the prior section, it can
be inferred that the extended model created with this deformation
function maintains the regularity of the Bardeen metric as given in
Eq.\eqref{27}. Figure \textbf{6} shows that the resulting spacetime
lacks asymptotic flatness.
\begin{figure}\center
\epsfig{file=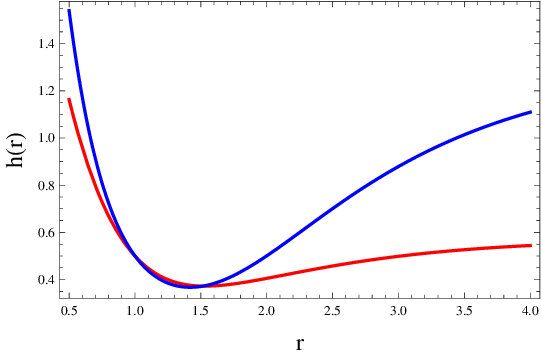,width=0.6\linewidth}\caption{Graphs of deformation
function $h(r)$ against $r$ with $\xi=0.2$ (red), $0.6$ (blue) for
model II.}
\end{figure}
\begin{figure}\center
\epsfig{file=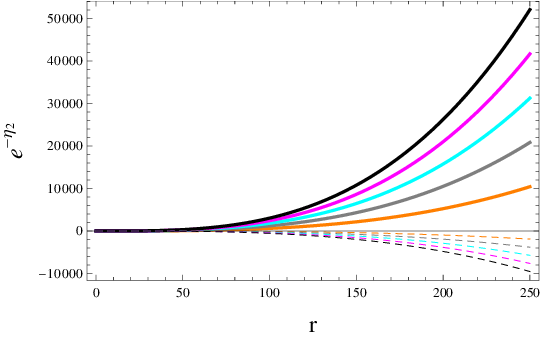,width=0.6\linewidth}\caption{Graph of $e^{-\eta_2}$
against $r$ with $\xi=0.2$ (solid), $0.6$ (dashed), $\sigma=0.2$
(orange), $0.4$ (gray), $0.6$ (cyan), $0.8$ (magenta) and $1$
(black) for model II.}
\end{figure}
\begin{figure}\center
\epsfig{file=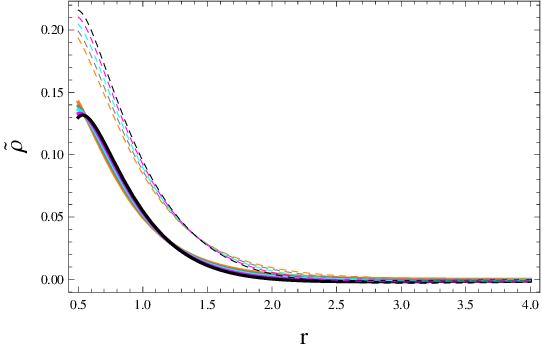,width=0.475\linewidth}
\epsfig{file=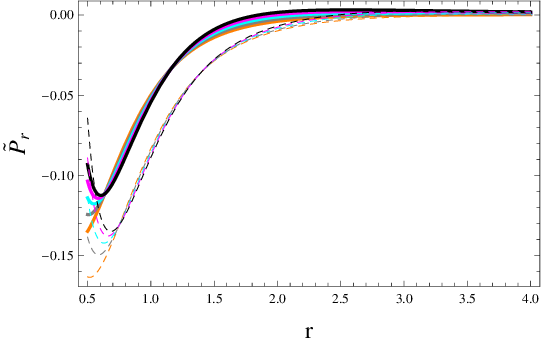,width=0.475\linewidth}
\epsfig{file=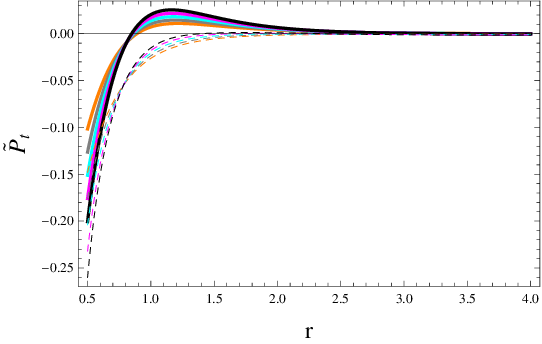,width=0.475\linewidth}\caption{Graphs of effective
parameters against $r$ with $\xi=0.2$ (solid), $0.6$ (dashed),
$\sigma=0.2$ (orange), $0.4$ (gray), $0.6$ (cyan), $0.8$ (magenta)
and $1$ (black) for model II.}
\end{figure}

The effective variables in Figure \textbf{7} provide further
insights into the derived model. Similar to the model discussed in
the previous section, this model features a positive density and a
negative radial pressure, while the tangential pressure alternates
between negative and positive values. With respect to the Rastall
parameter, the density varies directly, while the radial and
tangential pressures vary inversely. Regarding the decoupling
parameter $\sigma$, the effective parameters show a diverging
variation as they vary both directly and indirectly across various
intervals within their domains. Finally, we plot the energy
conditions in Figure \textbf{8} where a violation of some energy
conditions indicate an exotic source.
\begin{figure}\center
\epsfig{file=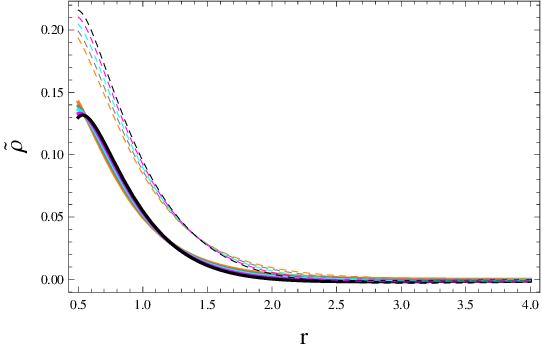,width=0.475\linewidth}
\epsfig{file=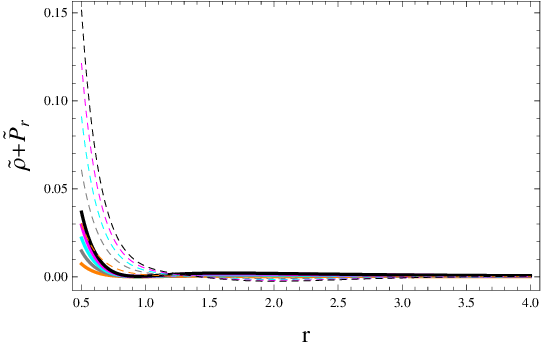,width=0.475\linewidth}
\epsfig{file=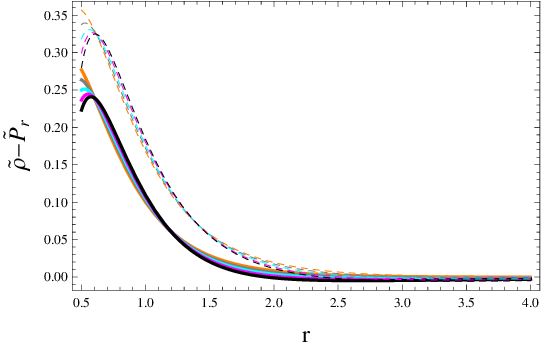,width=0.475\linewidth}
\epsfig{file=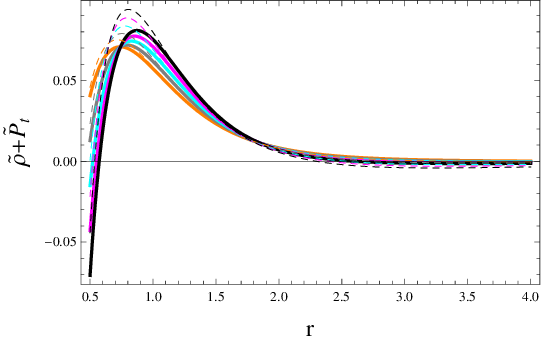,width=0.475\linewidth}
\epsfig{file=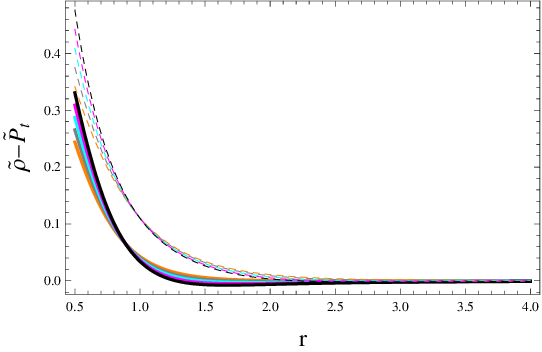,width=0.475\linewidth}
\epsfig{file=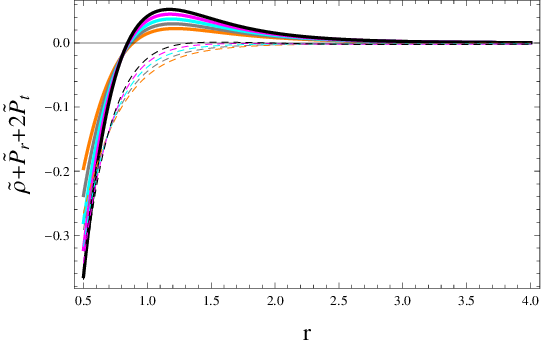,width=0.475\linewidth} \caption{Graphs of
effective parameters against $r$ with $\xi=0.2$ (solid), $0.6$
(dashed), $\sigma=0.2$ (orange), $0.4$ (gray), $0.6$ (cyan), $0.8$
(magenta) and $1$ (black) for model II.}
\end{figure}

\section{Some Thermodynamic Properties}

In this part, we explore various thermodynamic properties of black
holes. Factors such as temperature, entropy, and specific heat,
connect quantum mechanics, GR, and statistical mechanics. Studying
the radiation absorption and emission by black holes helps
scientists to discover details about the nature of spacetime, the
behavior of quantum fields in strong gravitational environments, and
the fundamental principles governing the fate of the universe. This
interdisciplinary approach not only deepens our comprehension of
black holes but also enhances our knowledge of the most extreme
physical laws.

\subsection{Hawking Radiation Temperature}

The temperature associated with Hawking radiation, often referred to
as Hawking temperature $T_H$, is a crucial thermodynamic
characteristic of black holes that provides significant insights
into the interplay between quantum mechanics and gravity. This
phenomenon, predicted by Stephen Hawking \cite{18}, involves the
emission of radiation from black holes as a result of quantum
effects occurring near the event horizon. This radiation temperature
is inversely related to the mass of the black hole, meaning that
smaller black holes emit more radiation and have higher
temperatures. This discovery not only challenges the traditional
view of black holes as entities that only absorb matter but also
indicates that they can gradually loose mass and energy, potentially
leading to their complete evaporation. Investigating Hawking
temperature offers a distinctive perspective on the quantum
properties of black holes, linking theoretical models with
observable cosmic events. The expression for the Hawking temperature
is
\begin{equation}\label{36}
T_H=\frac{k}{2\pi}=\frac{1}{4\pi}\bigg
|\frac{g_{tt,r}}{\sqrt{-g_{tt}g_{rr}}}\bigg |_{r=r_H}.
\end{equation}
\begin{figure}\center
\epsfig{file=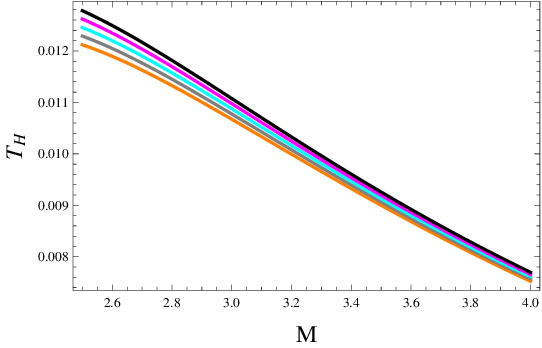,width=0.5\linewidth}\caption{Graph of $T_H$ against
$M$ with $\xi=0.2$ (solid), $0.6$ (dashed), $\sigma=0.2$ (orange),
$0.4$ (gray), $0.6$ (cyan), $0.8$ (magenta) and $1$ (black) for
model I.}
\end{figure}
\begin{figure}\center
\epsfig{file=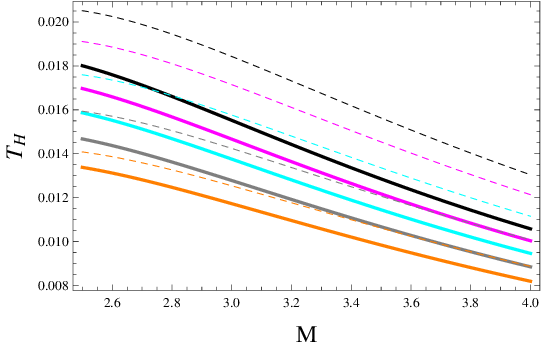,width=0.5\linewidth}\caption{Graph of $T_H$ against
$M$ with $\xi=0.2$ (solid), $0.6$ (dashed), $\sigma=0.2$ (orange),
$0.4$ (gray), $0.6$ (cyan), $0.8$ (magenta) and $1$ (black) for
model II.}
\end{figure}

The Hawking temperature for model I in Figure \textbf{9}
demonstrates appropriate behavior, indicating an inverse
relationship between temperature and black hole mass. However,
variations in the Rastall parameter show a negligible effect on the
Hawking temperature, whereas the decoupling parameter shows a direct
correlation. Similarly, the Hawking temperature for model II (Figure
\textbf{10}) exhibits acceptable behavior. In this case, contrary to
model I, the Rastall parameter has a noticeable impact, directly
influencing the Hawking temperature, alongside the decoupling
parameter.

\subsection{Specific Heat}

The specific heat is a critical thermodynamic parameter for
evaluating the thermal stability of black holes, measuring the
amount of heat needed to produce a small temperature change in a
black hole. In black hole thermodynamics, specific heat can indicate
stability properties. A positive specific heat means the black hole
can reach thermal equilibrium with its surroundings, implying
stability. Conversely, a negative specific heat signals instability,
leading to runaway heating or cooling during heat exchange. This
behavior is particularly evident in various black hole models, such
as Schwarzschild and Kerr black holes, where variations in specific
heat can highlight phase transitions or critical points in their
thermodynamic properties.
\begin{figure}\center
\epsfig{file=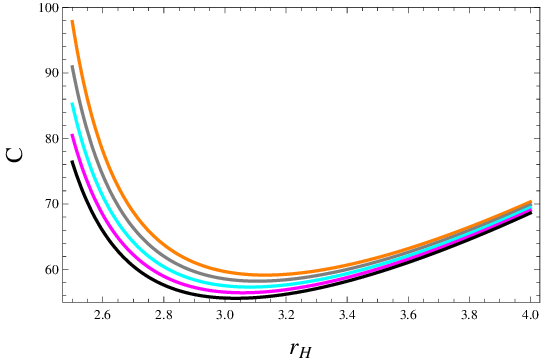,width=0.5\linewidth}\caption{Graph of $C$ against
$r_H$ with $\xi=0.2$ (solid), $0.6$ (dashed), $\sigma=0.2$ (orange),
$0.4$ (gray), $0.6$ (cyan), $0.8$ (magenta) and $1$ (black) for
model I.}
\end{figure}

The expression for the specific heat capacity is given by
\begin{equation}\label{37}
C=T_H\bigg(\frac{\partial S}{\partial T_H}\bigg)\bigg
|_{r=r_H}=T_H\bigg(\frac{\partial S}{\partial
r_H}\bigg)\bigg(\frac{\partial T_H}{\partial r_H}\bigg)^{-1},
\end{equation}
where
\begin{equation}\label{38}
S=\frac{1}{4}\int_0^{2\pi}\int_0^\pi\sqrt{g_{\theta\theta}
g_{\phi\phi}}d\theta d\phi=\pi r_H^2,
\end{equation}
denotes the Bekenstein-Hawking entropy \cite{49}. Figure \textbf{11}
shows a positive specific heat for the first model, in the interval
$2.5\leq r_H\leq 4$. This implies that the system is stable in this
interval. It is observed that the Rastall parameter shows no
variation with respect to the specific heat, while the decoupling
parameter varies inversely. For the second model (Figure
\textbf{12}), we observe a direct variation of the specific heat to
the Rastall parameter. The decoupling parameter varies inversely to
the specific heat, when considered with the lower value of the
Rastall parameter ($\xi=0.2$). The decoupling parameter, however,
varies directly with the specific heat corresponding to $(\xi=0.6)$.
This model also exhibits stability in the interval $2.5\leq r_H\leq
4$, as a positive specific heat is registered.
\begin{figure}\center
\epsfig{file=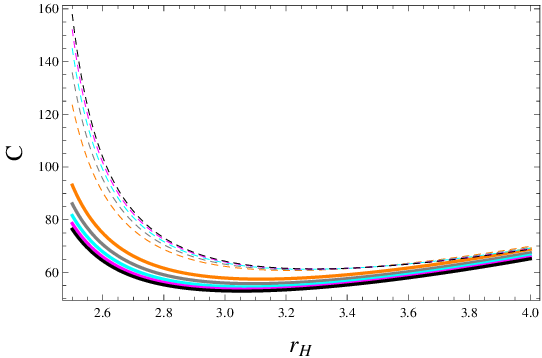,width=0.5\linewidth}\caption{Graph of $C$ against
$r_H$ with $\xi=0.2$ (solid), $0.6$ (dashed), $\sigma=0.2$ (orange),
$0.4$ (gray), $0.6$ (cyan), $0.8$ (magenta) and $1$ (black) for
model II.}
\end{figure}

\subsection{Hessian Matrix}

The Hessian matrix, particularly through its trace, provides
valuable insights into the thermodynamic stability of black holes.
This matrix comprises second-order partial derivatives of the
Helmholtz free energy, $F=E-ST_H$, with $E$ representing internal
energy, $S$ representing entropy, and $T_H$ representing the Hawking
temperature of the black hole. These derivatives are calculated with
respect to temperature and volume, where the temperature is
specified by the Hawking temperature and the volume is given by
$V=\frac{4}{3}\pi r^3_H$. The Hessian matrix is given by
\begin{equation}
H=\begin{pmatrix}
H_{11} & H_{12} \\
\\
H_{21} & H_{22}
\end{pmatrix}
=
\begin{pmatrix}
\frac{\partial^2F}{\partial T_H^2} & \frac{\partial^2F}{\partial T_H\partial V} \\
\\
\frac{\partial^2F}{\partial V\partial T_H} &
\frac{\partial^2F}{\partial V^2}
\end{pmatrix}.
\end{equation}
It can be observed that $det(H)=0$, which suggests that the given
Hessian matrix has an eigenvalue of zero. The positive definiteness
of the Hessian matrix cannot thus be exploited to determine the
stability of the system. We therefore determine the stability by
using the trace of the Hessian matrix given by
\begin{equation}
Tr(H)=H_{11}+H_{22}.
\end{equation}
Here, the criterion for stability is that $Tr(H)\geq 0$ \cite{50}.
Due to very lengthy and complicated expressions, the explicit form
of the trace is not written. However, the plots of the trace versus
horizon radius ($r_H$) are shown for both models.
\begin{figure}\center
\epsfig{file=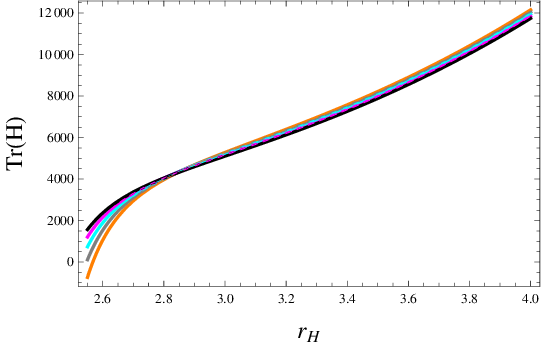,width=0.5\linewidth}\caption{Graph of $Tr(H)$
against $r_H$ with $\xi=0.2$ (solid), $0.6$ (dashed), $\sigma=0.2$
(orange), $0.4$ (gray), $0.6$ (cyan), $0.8$ (magenta) and $1$
(black) for model I.}
\end{figure}

The graphical analysis of $Tr(H)$ for the first model is shown in
Figure \textbf{13}, which indicates that the model is stable in the
interval $2.575\leq r_H\leq 4$. However, we observe a
null/negligible effect in the fluctuation of the Rastall parameter,
while the decoupling parameter shows a direct variation to $Tr(H)$
in the interval $2.575\leq r_H\leq 2.8$ and inverse variation in the
rest of the interval. For the second model (Figure \textbf{14}), we
deduce that the system is stable in the interval $2.5<r_H\leq 4$,
when considered with the Rastall parameter $\xi=0.2$. With the
Rastall parameter $\xi=0.6$, the interval of stability becomes
$2.65<r_H\leq 4$. It is thus deduced that for both Rastall
parameters used, the system is stable in the interval $2.65<r_H\leq
4$. Additionally, we observe an overall indirect variation of the
Rastall and decoupling parameters to the trace, $Tr(H)$.
\begin{figure}\center
\epsfig{file=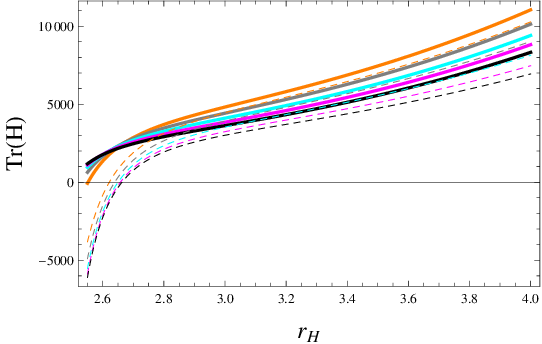,width=0.5\linewidth}\caption{Graph of $Tr(H)$
against $r_H$ with $\xi=0.2$ (solid), $0.6$ (dashed), $\sigma=0.2$
(orange), $0.4$ (gray), $0.6$ (cyan), $0.8$ (magenta) and $1$
(black) for model II.}
\end{figure}

\section{Conclusions}

In this study, we have identified minimally decoupled regular
Bardeen black hole solutions within the framework of Rastall
gravity. We have formulated the field equations and utilized the MGD
scheme to optimize the gravitational decoupling process. This method
separates the field equations into two distinct sets. The first set
corresponds to the regular Bardeen black hole, while the second set
involves an additional source, $\chi_{\upsilon\omega}$, which is
gravitationally linked to the primary source through the decoupling
parameter, $\sigma$. This supplementary source enables the extension
of the regular Bardeen black hole, allowing the derivation of new
black hole solutions that retain the physical properties of the
original Bardeen black hole.

The comprehensive solutions are derived through a linear combination
of the solutions to the subfield equations, which are obtained
post-decoupling. The solution to the first system is determined by
the metric components of the regular Bardeen black hole, whereas the
solution to the second system is derived using specific constraints
provided by a linear EoS. We have identified two extended solutions
corresponding to two particular instances of the given EoS. Our
findings indicate that both extended models maintain the regularity
of the original Bardeen black hole, aligning with the results in
reference \cite{51}. For both models, we have examined the effects
of the Rastall and decoupling parameters, $\xi$ and $\sigma$,
respectively, using the values $\xi=0.2,0.6$ and
$\sigma=0.2,0.4,0.6,0.8,1$.

We have observed a positive energy density paired with a negative
radial pressure for both models. Our investigation into the
asymptotic flatness of these models revealed that none of the
resulting solutions maintain asymptotic flatness. To analyze
asymptotic flatness, we have focused exclusively on the deformed
radial metric coefficient, as the temporal metric coefficient, which
remains unchanged, evidently approaches to 1 as $r$ becomes
arbitrarily large. Additionally, our findings indicate that both
models involve exotic matter, as they violate certain energy
conditions.

Our investigation into the thermodynamic properties of both models
revealed that black holes with lower mass emitted higher levels of
radiation. This outcome aligns with theoretical expectations, as the
emission of radiation results in evaporation and thus a reduction in
mass. Specifically, in the first model, the Rastall parameter
demonstrated only a minor fluctuation, whereas the decoupling
parameter has shown a direct correlation with the Hawking
temperature. Conversely, in the second model, both the Rastall and
decoupling parameters have displayed a direct relationship with the
Hawking radiation temperature.

We have examined thermodynamic stability of both models by analyzing
heat capacity and trace of the Hessian matrix. The heat capacity
analysis indicates that both models remain stable within the range
$2.5 \leq r_H \leq 4$. In contrast, the Hessian matrix trace
analysis reveals stability ranges of $2.575 \leq r_H \leq 4$ for the
first model and $2.65 < r_H \leq 4$ for the second model.
Consequently, we can conclude that both models exhibit stability
within the interval $2.65 < r_H \leq 4$.\\\\
\textbf{Data Availability:} No data was used for the research
described in this paper.

\end{document}